%% file: main_fusion.tex
\documentclass[conference]{IEEEtran}
\IEEEoverridecommandlockouts

\usepackage{cite}
\usepackage{amssymb}
\usepackage{textcomp}
\usepackage{graphicx, amsmath, amsfonts, bm, algorithm, algorithmic, subcaption, multirow, color}
\def\BibTeX{{\rm B\kern-.05em{\sc i\kern-.025em b}\kern-.08em
    T\kern-.1667em\lower.7ex\hbox{E}\kern-.125emX}}
\begin{document}

\title{Incorporating the ChEES Criterion into Sequential Monte Carlo Samplers\\
{\footnotesize 
}}

\author{Andrew Millard$^{\star}$, Joshua Murphy$^{\star}$, Daniel Frisch$^{\dagger}$, Simon Maskell$^{\star}$\\
{\normalsize{}$^{\star}$Dept. of EEE,
University of Liverpool, United Kingdom}\\
{\normalsize{}$^{\dagger}$Intelligent Sensor-Actuator-Systems Laboratory (ISAS), Karlsruhe Institute of Technology, Germany}\\
 
{\normalsize{}Emails: amill212@liverpool.ac.uk, jmurph98@liverpool.ac.uk, daniel.frisch@kit.edu, smaskell@liverpool.ac.uk}}

\maketitle

\input{0_abstract}
\input{1_intro}
\input{2_chees}

\input{3_smc}
\input{4_experiments}
\input{5_discussion}

\input{6_conc}

\section*{Acknowledgment}
AM and JM were funded by a Research Studentship jointly
funded by the EPSRC Centre for Doctoral Training in Distributed Algorithms EP/S023445/1. This work was funded by Dstl in collaboration with the Royal Academy of Engineering via "Dstl-RAEng Research Chair in Information Fusion" under task RQ0000040616. SM thanks Dstl, UK MOD and the Royal Academy of Engineering for supporting this work. The views and conclusions contained in this paper are of the authors and should not be interpreted as representing the official policies, either expressed or implied, of the UK MOD or the UK Government.

\bibliographystyle{ieeetr}
\bibliography{mybibliography}

\end{document}

%% file: 0_abstract.tex
\begin{abstract}
Markov chain Monte Carlo (MCMC) methods are a powerful but computationally expensive way of performing non-parametric Bayesian inference. MCMC proposals which utilise gradients, such as Hamiltonian Monte Carlo (HMC), can better explore the parameter space of interest if the additional hyper-parameters are chosen well. The No-U-Turn Sampler (NUTS) is a variant of HMC which is extremely effective at selecting these hyper-parameters but is slow to run and is not suited to GPU architectures. An alternative to NUTS, Change
in the Estimator of the Expected Square HMC (ChEES-HMC) was shown not only to run faster than NUTS on GPU but also sample from posteriors more efficiently. Sequential Monte Carlo (SMC) samplers are another sampling method which instead output weighted samples from the posterior. They are very amenable to parallelisation and therefore being run on GPUs while having additional flexibility in their choice of proposal over MCMC. We incorporate (ChEES-HMC) as a proposal into SMC samplers and demonstrate competitive but faster performance than NUTS on a number of tasks. 

\end{abstract}

%% file: 1_intro.tex
\section{Introduction}
Bayesian inference is a versatile way of making predictions and quantifying uncertainty while incorporating prior knowledge for a variety of applications such as deep learning \cite{neal2012bayesian}, epidemiology \cite{rasmussen2011inference}, and environmental modelling \cite{papaefthymiou2008mcmc}.

Inference on more complex posteriors may require the use of sampling methods such as Markov chain Monte Carlo (MCMC) \cite{andrieu2003introduction}. These sampling methods propose local moves within the distribution and build up a chain of particles which can be used to calculate statistics of functions on the distribution. MCMC can require many iterations to properly converge to a posterior so improved computational resources and tailoring the algorithm to exploit these architectures, often through parallelization,  have contributed to its increased popularity \cite{robert2018accelerating,mahani2015simd,mingas2017particle}. Better proposals such as Hamiltonian Monte Carlo (HMC) make MCMC more efficient by using gradients to inform local moves but introduce tunable hyper-parameters which can be difficult to select \cite{neal2012mcmc}. The most popular HMC variant is the No-U-Turn Sampler (NUTS) which automatically tunes these hyper-parameters \cite{hoffman2014no}. NUTS is the primary choice of sampler in numerous probabilistic programming languages like Stan, TensorFlow Probability and Numpyro \cite{carpenter2017stan,dillon2017tensorflowdistributions,phan2019composableeffectsflexibleaccelerated}. However, the NUTS algorithm is not well-suited to take advantage of GPUs due to its complex control flow and inherent recursion \cite{radul2020automaticallybatchingcontrolintensiveprograms,lao2020tfpmcmcmodernmarkovchain}.  Change in the Estimator of the Expected Square HMC (ChEES-HMC) was proposed as an alternative adaptive HMC variant better suited to GPU architectures and demonstrated significant speed-up while matching the performance of NUTS \cite{hoffman2021adaptive}.

Sequential Monte Carlo (SMC) samplers are another method of sampling from complex posteriors by again using local moves to iteratively propagate a set of weighted samples around a distribution \cite{del2006sequential}. The importance sampling components of SMC samplers are easy to parallelise and mitigate some of the concerns of running a number of MCMC chains in parallel and combining them when they may not all have converged \cite{drousiotis2025massively,varsi2019singlesmcsamplermpi}.

Our contribution is to incorporate ChEES into the SMC framework and explore the effect of the choice of the quasi-random number generator used to jitter trajectory length.

The paper proceeds as follows: Section \ref{sec:MCMC} gives an introduction to ChEES-HMC in the context of MCMC and Section \ref{sec:SMC} presents SMC and how ChEES is incorporated as a proposal. Section \ref{sec:experiments} lays out the experiments with Section \ref{sec:discussion} providing results and discussion with conclusions and future work given in Section \ref{sec:conclusion}.

%% file: 2_chees.tex
\section{Markov Chain Monte Carlo}\label{sec:MCMC}
Consider the problem of sampling parameters $\bm{\theta} \in \mathbb{R}^D$ from a posterior distribution proportional to $\pi(\bm{\theta}) = p(\mathbf{\mathbf{x}|\bm{\theta}})q_0(\bm{\theta})$ where $p(\mathbf{\mathbf{x}|\bm{\theta}})$ is a likelihood function and $q_0(\bm{\theta})$ is a specified prior distribution over the parameters.

MCMC is a common general-purpose way of obtaining samples from intractable posteriors. The algorithm proceeds with a user-selected initial state, $\bm{\theta}_0$ and subsequently building up a chain of $M$ parameter samples by proposing new samples according to a proposal distribution
\begin{equation}
    \bm{\theta}' \sim q(\cdot|\bm{\theta}_{m-1}).
\end{equation}

To ensure that samples come from the posterior distribution of interest, the sampler must be aperiodic, irreducible, obey detailed balance and the chain must leave the target distribution invariant. If the proposal is reversible, we can use the Metropolis-Hastings acceptance criterion and thus the samples come from the target \cite{hastings1970sampling} as our sampling process is now invariant. 
\begin{equation}
    \alpha(\bm{\theta}_{m-1}, \bm{\theta}') = \min \left( 1, \frac{\pi(\bm{\theta}') q(\bm{\theta}_{m-1} | \bm{\theta}')}{\pi(\bm{\theta}_{m-1}) q(\bm{\theta}' | \bm{\theta}_{m-1})} \right).
    \label{eq:mh_ratio}
\end{equation}

A newly proposed sample, $\bm{\theta}'$ is accepted and added to the chain if a random variable, $u \sim \mathcal{U}(0, 1)$, drawn from a uniform distribution is less than the Metropolis-Hastings criterion in \eqref{eq:mh_ratio}
\begin{equation}
    \bm{\theta}_m = 
    \begin{cases} 
    \bm{\theta}' & \text{if } u < \alpha(\bm{\theta}_{m-1}, \bm{\theta}'), \\
    \bm{\theta}_{m-1} & \text{otherwise}.
    \end{cases}
    \label{eq:mh_ratio_rn}
\end{equation}

\subsection{Hamiltonian Monte Carlo}
HMC is a sampler which uses gradients to make more informed moves around the posterior \cite{betancourt2017conceptual}. It augments the posterior to be $\pi(\bm{\theta}, \mathbf{p})$ with a momentum variable usually taken from a Gaussian distribution, $\bm{p} \sim \mathcal{N}(0, \mathcal{M})$, where $\mathcal{M}$ is the mass matrix and is typically set to an identity matrix $\mathcal{M} = \mathbf{I}_{D}$. The joint distribution can be written as
\begin{equation}
    \pi(\bm{\theta}, \mathbf{p}) = \exp\{-H(\bm{\theta}, \mathbf{p})\}, \label{eq:HMC_mh}
\end{equation}

\noindent where, $H(\cdot,\cdot)$, the Hamiltonian is

\begin{align}
    H(\bm{\theta}, \mathbf{p}) &= -\log\pi(\mathbf{p}|\bm{\theta})-\log \pi(\bm{\theta})\\
    &= -\frac{1}{2} \|\bm{p}\|^2-\log \pi(\bm{\theta}).
\end{align}

\noindent which can be interpreted as a combination of kinetic and potential energy respectively. Therefore, the sample location and momentum can be updated using Hamilton's equations. These equations are generally intractable so a more practical approach is to use the leapfrog integrator to solve them numerically 
\begin{align}
    \bm{p}_{l+\frac{1}{2}} &= \bm{p}_l + \frac{\epsilon}{2} \nabla \log \pi(\bm{\theta}_l)\label{equ:LF_p1/2}, \\
    \bm{\theta}_{l+1} &= \bm{\theta}_l + \epsilon \bm{p}_{l+\frac{1}{2}}\label{equ:LF_theta}, \\
    \bm{p}_{l+1} &= \bm{p}_{l+\frac{1}{2}} + \frac{\epsilon}{2} \nabla \log \pi(\bm{\theta}_{l+1}).
    \label{equ:LF_p}
\end{align}

The leapfrog process is repeated for a certain number of user-specified leapfrog steps $L$ with step size $\epsilon$. Pseudocode for the leapfrog algorithm can be found in Algorithm \ref{alg:LF_example}. Upon completion of $L$ steps, leapfrog proposes a new sample and momentum, setting $\bm{\theta}'=\bm{\theta}_L$ and $\mathbf{p}'=\mathbf{p}_L$, which are accepted according to the criterion
\begin{align}
    \alpha((\bm{\theta}_{m-1}, \mathbf{p}_{m-1}), (\bm{\theta}', \mathbf{p}')) = \label{eq:HMC_accept_reject}\\
    \min(1, \exp(-H(\bm{\theta}', \mathbf{p}') + H(\bm{\theta}_{m-1}, \mathbf{p}_{m-1})).  \nonumber
\end{align}

\begin{algorithm}[H]
\caption{Leapfrog Algorithm}\label{alg:LF_example}
\begin{algorithmic}
\REQUIRE Initial state $\bm{\theta}$, momentum $\mathbf{p}$, step size $\epsilon$ and number of leapfrog steps $L$
\FOR{$l = 1$ to $L$}
    \STATE Half step update the momentum \eqref{equ:LF_p1/2}
    \STATE Update the state \eqref{equ:LF_theta}
    \STATE Complete the momentum update \eqref{equ:LF_p}
\ENDFOR
\STATE \textbf{return} $\bm{\theta}_L, \mathbf{p}_L$
\end{algorithmic}
\end{algorithm}

The step size $\epsilon$ can be selected by numerous adaptive schemes such as dual-averaging \cite{nesterov2009primal}, optimisation of the expected squared jump distance \cite{pasarica2010adaptively,wang2013adaptive} and other adaptive MCMC methods. These do not generally interfere with the benefits of SMC and its parallelisation capacities \cite{buchholz2021adaptive}. However, selection of $L$  is a little more challenging.

\subsection{No-U-Turn Sampler}
The most popular adaptive trajectory length selection algorithm is NUTS \cite{hoffman2014no}. At each MCMC iteration, NUTS builds up a binary tree, using the leapfrog algorithm to build up trajectories in both directions alternately while doubling the number of leapfrog steps each time it considers switching direction. The construction of the tree stops when the trajectory makes a "U-turn",
\begin{equation}
    (\bm{\theta}^+-\bm{\theta}^-)\cdot \mathbf{p}^- < 0 \quad \text{or} \quad (\bm{\theta}^+-\bm{\theta}^-)\cdot \mathbf{p}^+ < 0,\label{eq:u-turn}
\end{equation}

\noindent so that the tree ends up proposing just enough samples that some are far away from the point from which the tree is grown from. $\bm{\theta}^+$ and $\bm{\theta}^-$ represent the two furthest left and right points respectively in the binary tree. The trajectory is then randomly sampled to give $\bm{\theta}'$ which is accepted or rejected based on the criterion in \eqref{eq:HMC_mh}. 

Additional steps are taken to maintain detailed balance, such as requiring that the points within the binary tree meet four key conditions in order to be selected as a potential sample $\bm{\theta}'$. The full details of these conditions can be found in the original paper \cite{hoffman2014no}. 

\subsection{Change in the Estimator of the Expected Square}
Using the Change in the Estimator of the Expected Square (ChEES) criterion \cite{hoffman2021adaptive} is an effective alternative for the NUTS algorithm when adapting the trajectory length. The ChEES criterion is the following 

\begin{equation}
    \textbf{ChEES} = \frac{1}{4} \mathbb{E}[(|| \bm{\theta'} - \mathbb{E}[\bm{\theta}]||^2 - ||\bm{\theta} - \mathbb{E}[\bm{\theta}]||^2)^2]
\end{equation}



This criterion promotes exploration by maximizing changes in sample variance, thereby reducing autocorrelation. For each chain \( c \in \{1, \dots, C\} \), it is evaluated using proposed and previous states \( \bm{\theta}^{(c)'} \), \( \bm{\theta}_{m-1}^{(c)} \), and their means \( \hat{\bm{\theta}}' \), \( \hat{\bm{\theta}} \). The gradient estimate \( \hat{g}^{(c)} \) reflects how trajectory length influences sample dispersion, scaled by the jittered length \( l_m \) and momentum \( \bm{p}^{(c)} \).

To enhance exploration, trajectory lengths are jittered via a Halton sequence \( h_m \), with \( l_m = h_m L_{m-1} \)~\cite{halton1960efficiency}. During warm-up, ChEES is maximized using gradient descent to adapt \( L_m \), based on the weighted average gradient:
\begin{equation}
    \hat{g} = \frac{\sum_{c} \alpha^{(c)} \hat{g}^{(c)}}{\sum_{c} \alpha^{(c)}}
\end{equation}
This updates \( \log L_m \) via Adam, with a moving average \( \bar{L} \) maintained. After \( m = M_{\text{warmup}} \), \( L_m \) is fixed to \( \bar{L} \).

Compared to NUTS, ChEES is often more efficient while still effectively adapting trajectory lengths. In SMC, multiple chains allow trajectory length to be tuned in parallel during warm-up. Algorithm \ref{alg:smc_ChEES} provides the full procedure.

\begin{algorithm}
    \caption{ChEES-HMC running on C chains.}
    \begin{algorithmic}[1]
    \label{alg:smc_ChEES}
    \REQUIRE Initial state $\bm{\theta}_0^{(c)}$ for each chain $c \in \{1,\dots, C\}$, step size $\epsilon$, initial trajectory length $L_0$, desired number of samples $M$, number of adaptation steps $M_{\text{warmup}}$, random number sequence $h_{1:M}$.
    \STATE Initialise moving averages $\bar{L} = 0$.
    \FOR{$m = 1$ to $M$}
        \STATE Sample momentum $\bm{p}^{(c)} \sim \mathcal{N}(0, \mathcal{M})$.
        \STATE Select jittered trajectory length $l_m = h_m L_{m-1}$.
        \STATE Propose new sample $\bm{\theta}^{(c)'}$ and momentum $\mathbf{p}^{(c)'}$ using $\text{leapfrog}(\bm{\theta}^{(c)}_{m-1}, \mathbf{p}^{(c)'}, \epsilon, \lceil l_m / \epsilon \rceil)$.
        \STATE Compute acceptance probabilities $\alpha^{(c)}$ using \eqref{eq:HMC_accept_reject} 
        \STATE Select $\bm{\theta}_m^{(c)}$ and $\mathbf{p}_m^{(c)}$ according to \eqref{eq:mh_ratio_rn}
        \IF{$m < M_{\text{warmup}}$}
            \STATE Estimate the mean of the proposed and old states:  
            \STATE \hspace{1cm} $\hat{\bm{\theta}}' = \frac{1}{C} \sum_{c} \bm{\theta}^{(c)'}, \quad 
            \hat{\bm{\theta}} = \frac{1}{C} \sum_{c} \bm{\theta}^{(c)}_{m-1}$
            \STATE Compute trajectory gradient estimates:  
            \STATE \hspace{1cm} $
            \begin{aligned}
                \hat{g}^{(c)} &= l_m \Big( \|\bm{\theta}^{(c)'} - \hat{\bm{\theta}}'\|^2 \\
                &\quad - \|\bm{\theta}^{(c)}_{m-1} - \hat{\bm{\theta}}\|^2 \Big) (\bm{\theta}^{(c)'} - \hat{\bm{\theta}}')^\top \mathbf{p}^{(c)}
            \end{aligned}$
            \STATE Update log-trajectory length $\log L_m$ with Adam using weighted gradient:
            \STATE \hspace{1cm} $\hat{g} = \frac{\sum_{c} \alpha^{(c)} \hat{g}^{(c)}}{\sum_{c} \alpha^{(c)}}$
            \STATE Update moving averages trajectory $\bar{L} \gets 0.9 \bar{L} + 0.1 L_m$
        \ENDIF
        \IF{$m = M_{\text{warmup}}$}
            \STATE $L_{m:M} \gets \bar{L}$
        \ENDIF
    \ENDFOR
    \end{algorithmic}
\end{algorithm}

%% file: 3_smc.tex
\section{Sequential Monte Carlo}\label{sec:SMC}
SMC is another algorithm for targeting static posterior distributions via $K$ sequential importance sampling steps and resampling when necessary \cite{del2006sequential}. The joint distribution of all states until $k=K$ is defined as
 \begin{equation}
        \pi(\bm{\theta}_{1:K}) = \pi(\bm{\theta}_{K}) \prod_{k=1}^{K} L(\bm{\theta}_{k-1} | \bm{\theta}_{k}),
    \end{equation}
    
\noindent where $L(\bm{\theta}_{k-1} | \bm{\theta}_{k})$ is the L-kernel, which is a user-defined probability distribution. The choice of this distribution can greatly impact the efficacy of the sampler \cite{green2022increasing}.

At $k=1$, $J$ samples $j = 1, \dots, J$ are drawn from a prior distribution $q_0(\cdot)$ as follows:
\begin{equation} \label{eq: init_sample}
        \bm{\theta}^j_0 \sim q_0(\cdot), \ \ \forall j,
    \end{equation}
    and weighted according to
    \begin{equation} 
        \mathbf{w}^j_1 = \frac{\pi(\bm{\theta}^j_0)}{q_0(\bm{\theta}^j_0)}, \ \ \forall j\label{eq:init_weights}.
    \end{equation}
At $k>1$, subsequent samples are proposed based on samples from the previous iteration via a proposal distribution, $q(\bm{\theta}^j_k|\bm{\theta}^j_{k-1})$ by 
\begin{equation} \label{eq:proposal}
        \bm{\theta}^j_k \sim q(\cdot|\bm{\theta}^j_{k-1}).
\end{equation}
    
These samples are then weighted according to 
    \begin{equation}
        \mathbf{w}^j_{k} = \mathbf{w}^j_{k-1} \frac{\pi(\bm{\theta}^j_{k})}{\pi(\bm{\theta}^j_{k-1})} \frac{L(\bm{\theta}^j_{k-1}|\bm{\theta}^j_{k})}{q(\bm{\theta}^j_{k}|\bm{\theta}^j_{k-1})}, \ \ \forall j.\label{eq:l_weights}
    \end{equation}
    
SMC samplers compute the Effective Sample Size (ESS) as a measure of the efficiency of the sampler at iteration $k$ by
\begin{equation} \label{eq:ess_smc}
        J^\text{eff} = \frac{1}{\sum\nolimits_{j=1}^{{J}} \left(\tilde{\mathbf{w}}_{k}^{j}\right)^2},
\end{equation}

\noindent using the sum of the normalised weights which are calculated from
\begin{equation}
     \tilde{\mathbf{w}}_{k}^{j} = \frac{\mathbf{w}_{k}^{j}}{\sum\nolimits_{j=1}^{N} \mathbf{w}_{k}^{j}}, \ \ \forall j.
        \label{eq: normalise_smc2}
    \end{equation} 
As iterations continue, one weight tends to dominate which is known as particle degeneracy and can be mitigated by resampling. Resampling is undertaken if $J^\text{eff}<J/2$. There are a variety of potential resampling schemes \cite{douc2005comparison} including the optimally parallelised systematic resampling schemes outlined in \cite{varsi2021log2,rosato2023logn}. Here we utilise multinomial resampling for ease of implementation. Samples are assigned an unnormalised weight of $\frac{1}{J}$ after resampling.

The weighted samples can be used to picture the whole distribution as well as realise estimates of the expectations of functions on the distribution through 
\begin{align}
    \mathbb{E}_{\pi}\left[ f(\bm{\theta}_k)\right] &= \int f(\bm{\theta}_k)\pi(\bm{\theta}_k)d\bm{\theta}_k \nonumber\\ 
    &= \int f(\bm{\theta}_k)\frac{\pi(\bm{\theta}_k)}{q(\bm{\theta}_k)} q(\bm{\theta}_k) d\bm{\theta}_k \approx \Tilde{\mathbf{f}}_k
\end{align}

\begin{equation} 
        \Tilde{\mathbf{f}}_{k} = \sum\nolimits_{j=1}^{J} \Tilde{\mathbf{w}}^j_{k} f(\bm{\theta}^j_{k}), \label{eq:realised_estimates}
\end{equation}

Pseudocode for a generic SMC sampler can be found in Algorithm \ref{alg:smc_BNN}.

\begin{algorithm}[tb]
   \caption{SMC sampler running for $K$ iterations and $J$ samples.}\label{alg:smc_BNN}
\begin{algorithmic}
    \STATE Sample $\{\bm{\theta}^{(j)}_0\}^J_{j=1} \sim q_0(\cdot)$
    \STATE Set initial weights $\mathbf{w}_0^{j}$ using \eqref{eq:init_weights}
    \FOR{$k=1$ {\bfseries to} $K$}
    \FOR{$j=1$ {\bfseries to} $J$}
    \STATE Normalise weights using \eqref{eq: normalise_smc2}
    \ENDFOR
    \STATE Calculate $J_\mathrm{eff}$ using \eqref{eq:ess_smc}
    \IF{$J_\mathrm{eff} < J/2$}
    \STATE Resample $[\bm{\theta}^{1}_k ... \bm{\theta}^{J}_k]$ with probability $[\Tilde{\mathbf{w}}^{1}_k ... \Tilde{\mathbf{w}}^{J}_k]$
    \STATE Reset all weights to $\frac{1}{J}$
    \ENDIF
    \FOR{$j=1$ {\bfseries to} $J$}
    \STATE Propagate samples $\bm{\theta}^j_{k-1}$ according to \eqref{eq:proposal}
    \STATE Update sample weights $\mathbf{w}_k^{j}$ using \eqref{eq:l_weights}
    \ENDFOR
    \ENDFOR
\end{algorithmic}
\end{algorithm}

\subsection{Proposals in SMC}
A common but naive choice of the proposal distribution in \eqref{eq:proposal} is a Gaussian with a mean of $\bm{\theta}^j_{k-1}$ and a covariance of $\bm{\Sigma} \in \mathbb{R}^{D\times D}$, such that
\begin{equation}      
q(\bm{\theta}^j_k|\bm{\theta}^j_{k-1}) = \mathcal{N}(\bm{\theta}^j_k; \bm{\theta}^j_{k-1}, \bm{\Sigma}), \ \ \forall j.
\end{equation}

This is also referred to as a random walk proposal. Gradient-based proposals originating from MCMC like Langevin \cite{sim2012information}, HMC \cite{gunawan2018robust,daviet2018inference} and NUTS \cite{devlin2024NUTS} have been effectively incorporated into SMC. MCMC proposals can be included in SMC with or without an accept-reject step \cite{devlin2024NUTS,buchholz2021adaptive}. Here we choose not to have an accept-reject step when including ChEES as a proposal in SMC (SMC-ChEES) and comparing to NUTS in SMC.

A sub-optimal but easily implementable approach to selecting the L-kernel in \eqref{eq:l_weights} is to choose the same distribution as the forwards proposal
\begin{equation}
        L(\bm{\theta}^j_{k-1}|\boldsymbol{\theta}^j_{k}) = q(\bm{\theta}^j_{k-1}|\bm{\theta}^j_{k}), \ \ \forall i,
    \end{equation}

With gradient-based proposals both proposal and L-kernel can be evaluated in terms of the stochastic momentum component $\mathbf{p}$ \cite{rosato2024enhanced,devlin2024NUTS}
\begin{align}
       q(\bm{\theta}^j_k|\bm{\theta}^j_{k-1})=
             \mathcal{N}(\bm{p}_{k-1}; 0,\mathcal{M}) \begin{vmatrix}
        \frac{df_{\text{LF}}(\bm{\theta}_{k-1}, \bm{p}_{k-1})}{d\bm{p}_{k-1}}
    \end{vmatrix}^{-1}, \label{eq:proposal_momentum} \\
     L(\bm{\theta}^j_{k-1}|\bm{\theta}^j_{k})= \mathcal{N}(-\bm{p}_k; 0,\mathcal{M}) \begin{vmatrix}
            \frac{df_{\text{LF}}(\bm{\theta}_k, -\bm{p}_k)}{d\bm{p}_k}
        \end{vmatrix}^{-1}. \label{eq:l-kernel_momentum}
\end{align} 

where $f_{\text{LF}}$ is the leapfrog process. When using \eqref{eq:proposal_momentum} and \eqref{eq:l-kernel_momentum} their Jacobians cancel in \eqref{eq:l_weights}. We also apply this change of variable to the proposal and L-kernel when we evaluate SMC-ChEES.

\subsection{SMC-ChEES}
SMC-ChEES is an extension of the HMC proposals discussed in section 2, as it provides a principled alternative trajectory adaption technique to NUTS. When using this in an SMC context, we can use $J$ random seeds to jitter each trajectory length by a different amount at each iteration, and then use this to compute the acceptance rate $\alpha$ needed for the optimization procedure to maximize ChEES during the warm up phase. Even after warm up, although we fix $L$ we can still use different jitters for each sample to effectively explore the distribution.  

\subsubsection{Random Number Generators (RNGs)}
In \cite{hoffman2021adaptive} the random number sequence $h_{1:M}$ was generated from a 1-dimensional Halton sequence \cite{halton1960efficiency}. We explore the usage of the following random and quasi-random number generation schemes to generate a matrix of $J \times K$ random numbers $h_{1:JK}$ for use with ChEES as a proposal in an SMC sampler. 

\begin{enumerate}
    \item No jitter: all $J \times K$ numbers are set to 1.
    \item Uniform random: the random number sequence is drawn from a standard uniform
        \begin{equation}
            h_{jk} \sim U(0,1).
        \end{equation}
    \item N-d Halton: 
    \begin{equation}
        h_{jk} = \sum_{d=0}^{\infty} \text{digits}^k_d(j) k^{-d-1} \label{eq:halton}
    \end{equation}
    where $\text{digits}_d(j)$ is the d$^{\text{th}}$ digit of $j$ represented in base-$k$ with the order of the digits reversed.
    \item N-d Inverse Halton: as in \eqref{eq:halton} but the matrix is then sorted in reverse order of $k$ to ensure lower discrepancy bases are used at the end of the sampling sequence.  
    \item 1-d Halton:  $J \times K$ numbers are taken from \eqref{eq:halton} with base set to 2. 
    \item N-d Primes:
    \begin{equation}
        h_{jk} = j{\sqrt{\mathbb{P}}_k} \mod 1\label{eq:primes}
    \end{equation}
    where $\mathbb{P}_k$ is the kth prime number. 
    \item N-d Inverse Primes: as in \eqref{eq:primes} with the matrix sorted in reverse order of $k$.
    \item 1-d Golden Ratio:
    \begin{equation}
         \frac{((j-1)K + k)(\sqrt{5}-1)}{2} \mod 1.
    \end{equation}
    \item Equidistant: $J$ points are created from $h_{j} = j/J$. These $h_j$ points are shuffled $K$ times to fill the $h_{jk}$ matrix.
    \item Offset Equidistant: as with Equidistant but each number is also perturbed with a draw from a $U(0,0.1)$. 
    \item N-d Sobol:
    \begin{equation}
        h_{jk} = \text{digits}_1^2(j)\nu_1^k \oplus \text{digits}_2^2(j)\nu_2^k \oplus \dots \label{eq:sobol}
    \end{equation}
    where $\nu_d^k$ are direction numbers typically obtained as the coefficients of a primitive polynomial. We use the scipy implementation which takes direction numbers from \cite{joe2008constructing}.
    \item Inverse N-d Sobol: as in \eqref{eq:sobol} with the matrix sorted in reverse order of $k$.
    \item 1-d Sobol: $J \times K$ numbers are taken from \eqref{eq:halton} with $k=1$.
\end{enumerate}

%% file: 4_experiments.tex
\section{Experimental Set-up}\label{sec:experiments}

\begin{table*}[!t]
    \centering 
\makebox[\textwidth]{\begin{tabular}{|c|cc|cc|cc|cc|}
        \hline
        & \multicolumn{2}{c|}{Gaussian} & \multicolumn{2}{c|}{Ill-conditioned Gaussian} & \multicolumn{2}{c|}{Banana} & \multicolumn{2}{c|}{German Credit} \\  
        & $\nabla\text{eval}/N$ & $J^{\text{eff}}/\nabla\text{eval}$ & $\nabla\text{eval}/N$ & $J^{\text{eff}}/\nabla\text{eval}$ & $\nabla\text{eval}/N$ & $J^{\text{eff}}/\nabla\text{eval}$ & $\nabla\text{eval}/N$ & $J^{\text{eff}}/\nabla\text{eval}$ \\
        \hline
        \textbf{NUTS} & 63.95 & 1.56e-02 & 1771.93 & 4.16e-04 & 468.58 & 2.12e-03 & 1952.19 & 3.61e-04 \\
\textbf{No Jitter} & 12.65 & 7.91e-02 & 501.00 & 1.63e-03 & 25.84 & 3.87e-02 & 501.00 & 1.40e-03 \\
\textbf{1-d Uniform} & 6.55 & 1.53e-01 & 501.00 & 1.27e-03 & 57.10 & 1.74e-02 & 143.48 & 5.11e-03 \\
\textbf{N-d Halton} & 11.02 & 9.08e-02 & 501.00 & 1.25e-03 & 73.80 & 1.35e-02 & 52.66 & 1.46e-02 \\
\textbf{N-d Inverse Halton} & 17.86 & 5.60e-02 & 501.00 & 1.21e-03 & 34.73 & 2.87e-02 & 72.77 & 1.04e-02 \\
\textbf{1-d Halton} & 8.01 & 1.25e-01 & 2.01 & 3.20e-01 & 21.96 & 4.55e-02 & 3.58 & 2.33e-01 \\
\textbf{N-d Primes} & 3.59 & 2.78e-01 & 501.00 & 1.20e-03 & 43.50 & 2.29e-02 & 102.15 & 7.50e-03 \\
\textbf{N-d Inverse Primes} & 5.55 & 1.80e-01 & 501.00 & 1.35e-03 & 50.34 & 1.98e-02 & 96.48 & 8.72e-03 \\
\textbf{1-d Golden Ratio} & 5.23 & 1.91e-01 & 501.00 & 1.26e-03 & 61.37 & 1.62e-02 & 92.92 & 9.00e-03 \\
\textbf{N-d Equidistant} & 6.46 & 1.55e-01 & 501.00 & 1.25e-03 & 52.61 & 1.89e-02 & 106.21 & 7.55e-03 \\
\textbf{N-d Offset Equidistant} & 9.06 & 1.10e-01 & 501.00 & 1.22e-03 & 52.63 & 1.89e-02 & 186.85 & 4.08e-03 \\
\textbf{N-d Sobol} & 7.28 & 1.37e-01 & 501.00 & 1.24e-03 & 62.26 & 1.60e-02 & 147.55 & 5.45e-03 \\
\textbf{N-d Inverse Sobol} & 8.68 & 1.15e-01 & 501.00 & 1.24e-03 & 77.89 & 1.28e-02 & 101.76 & 7.32e-03 \\
\textbf{1-D Sobol} & 8.76 & 1.14e-01 & 2.47 & 2.64e-01 & 34.43 & 2.90e-02 & 4.62 & 1.96e-01 \\
        \hline
    \end{tabular}}
    \caption{Number of gradient evaluations per sample (smaller is better) and effective sample size per gradient evaluation (larger is better) averaged across iterations.}
    \label{tab:grad_eval_ess}
\end{table*}

In this section we present the distributions we sample from for evaluation of our method. In all examples the SMC samplers have $J=1000$ particles and are run for $K=200$ iterations, the first 100 of which are taken as burn-in. Initial samples are drawn from a prior of appropriate dimensionality for the target $\bm{\theta}_0 \sim \mathcal{N}(\mathbf{0}_{D}, \mathbf{I}_{D})$. ChEES is initialised with $L=5$. Each sampler is run 10 times and an average of estimates and metrics is reported.

\subsection{Gaussian}
The first example is a multivariate Gaussian with $D=5$ and parameters
\begin{align}
    \bm{\theta} &\sim \mathcal{N}(\bm{\mu}, \bm{\Sigma}) \\
    \bm{\mu} &= [-4, -2, 0, 2, 4]^T \\
    \bm{\Sigma} &= \operatorname{diag}(1, 1.5, 2, 2.5, 3)
\end{align}
A step size of $\epsilon=0.1$ was utilised.

\subsection{Ill-conditioned Gauss}
Another multivariate Gaussian with $D=100$. A random orthogonal matrix $\mathbf{Q} \in \mathbb{R}^{100 \times 100}$ is drawn uniformly from the Haar measure on the orthogonal group $O(100)$, ensuring that $\mathbf{Q} \mathbf{Q}^\top = \mathbf{I}_{100}$. The eigenvalues $\{\lambda_i\}_{i=1}^{100}$ of the covariance matrix $\bm{\Sigma}$ are drawn from a Gamma distribution with shape parameter 0.5 and scale parameter 1. The condition number of $\bm{\Sigma}\approx 1.3 \times 10^5$ \cite{hoffman2021adaptive}. A step size of $\epsilon=0.001$ was used.

\begin{align}
    \bm{\theta} &\sim \mathcal{N}(\bm{\mu}, \bm{\Sigma}) \\
    \bm{\mu} &= [0, \dots, 0]^\top \in  \mathbb{R}^{100}\\
    \bm{\lambda}_j &\sim \text{Gamma}(0.5, 1), \quad j = 1, 2, \dots, 100 \\
    \mathbf{Q} &\sim \text{Haar}(O(100)) \\
    \bf{\Sigma} &= \mathbf{Q} \bm{\lambda} \mathbf{Q}^\top.
\end{align}
    
\subsection{Rosenbrock Distribution}
The Rosenbrock (banana) distribution with $D=2$. The joint distribution is a product of the marginals in dimension $\bm{\theta}_{(1)}$ and $\bm{\theta}_{(2)}$
\begin{align}
    \bm{\theta}_{(1)} &\sim \mathcal{N}(0, 10)\\
    \bm{\theta}_{(2)} &\sim \mathcal{N}(0.03(\bm{\theta}_{(1)}^2-100), 1)
\end{align}
A step size of $\epsilon=0.01$ was used.

\subsection{German Credit}
A logistic regression on the numerical German credit dataset \cite{statlog_(german_credit_data)_144}. Here $D=25$. A step size of $\epsilon=0.001$ was used.

\begin{align}
    \mathbf{y}_n &\sim \text{Bernoulli}(\sigma(\bm{\theta}^\top \mathbf{x}_n)) \\
    \sigma(x) &\overset{\Delta}{=} \frac{1}{1+e^{-x}}
\end{align}

%% file: 5_discussion.tex
\section{Results}\label{sec:discussion}

\begin{figure*}[!t]
    \begin{subfigure}{\textwidth}
        \includegraphics[width=\textwidth]{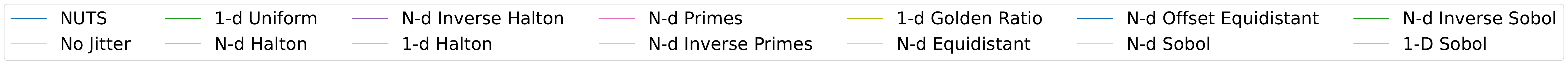}
    \end{subfigure}
    \centering
    \newcommand{\SubFigureWidth}{0.23}
    \begin{subfigure}{\SubFigureWidth\textwidth}
    \includegraphics[width=\textwidth]{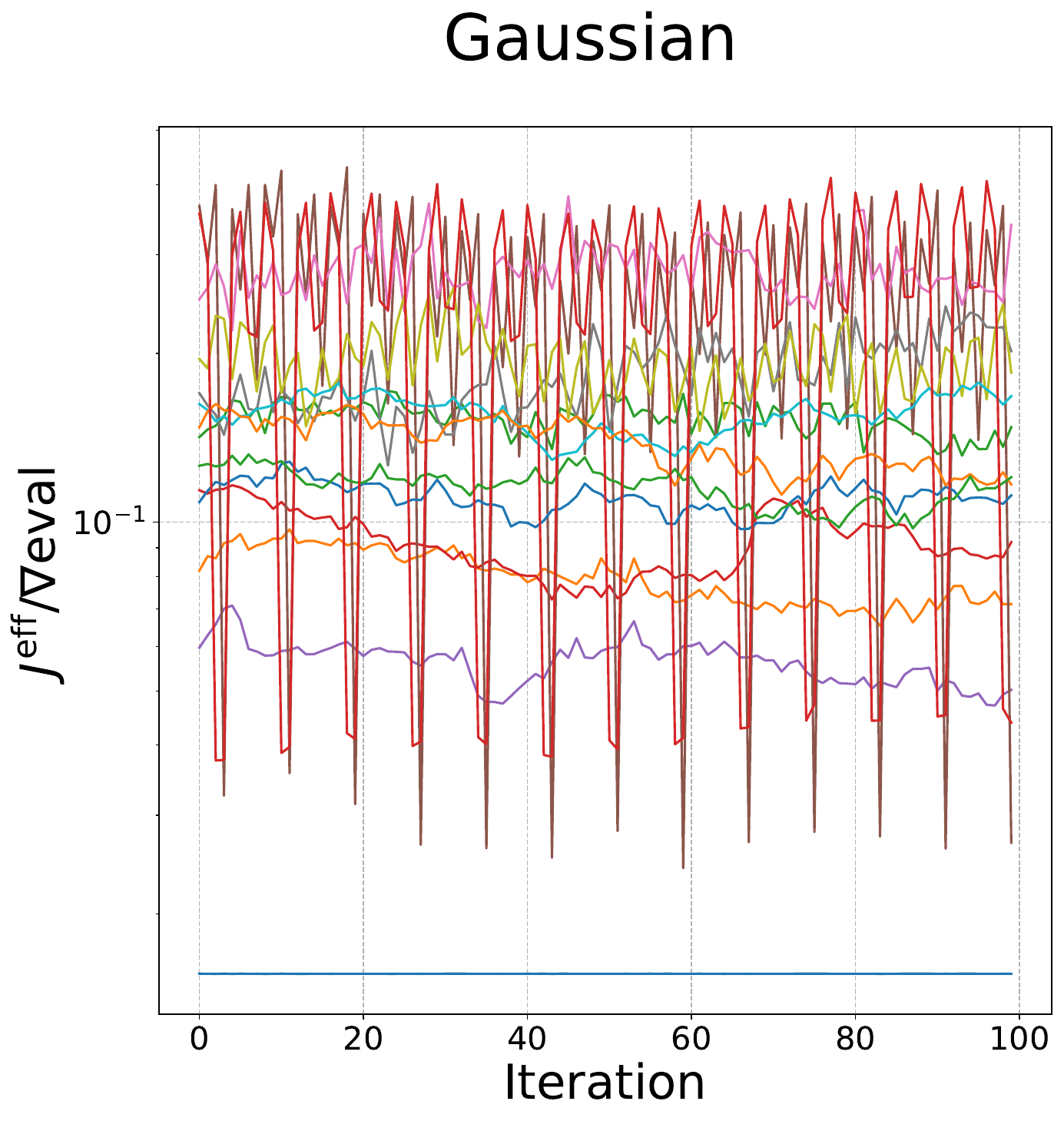}
    \end{subfigure}
    \hfill
    \begin{subfigure}{\SubFigureWidth\textwidth}
    \includegraphics[width=\textwidth]{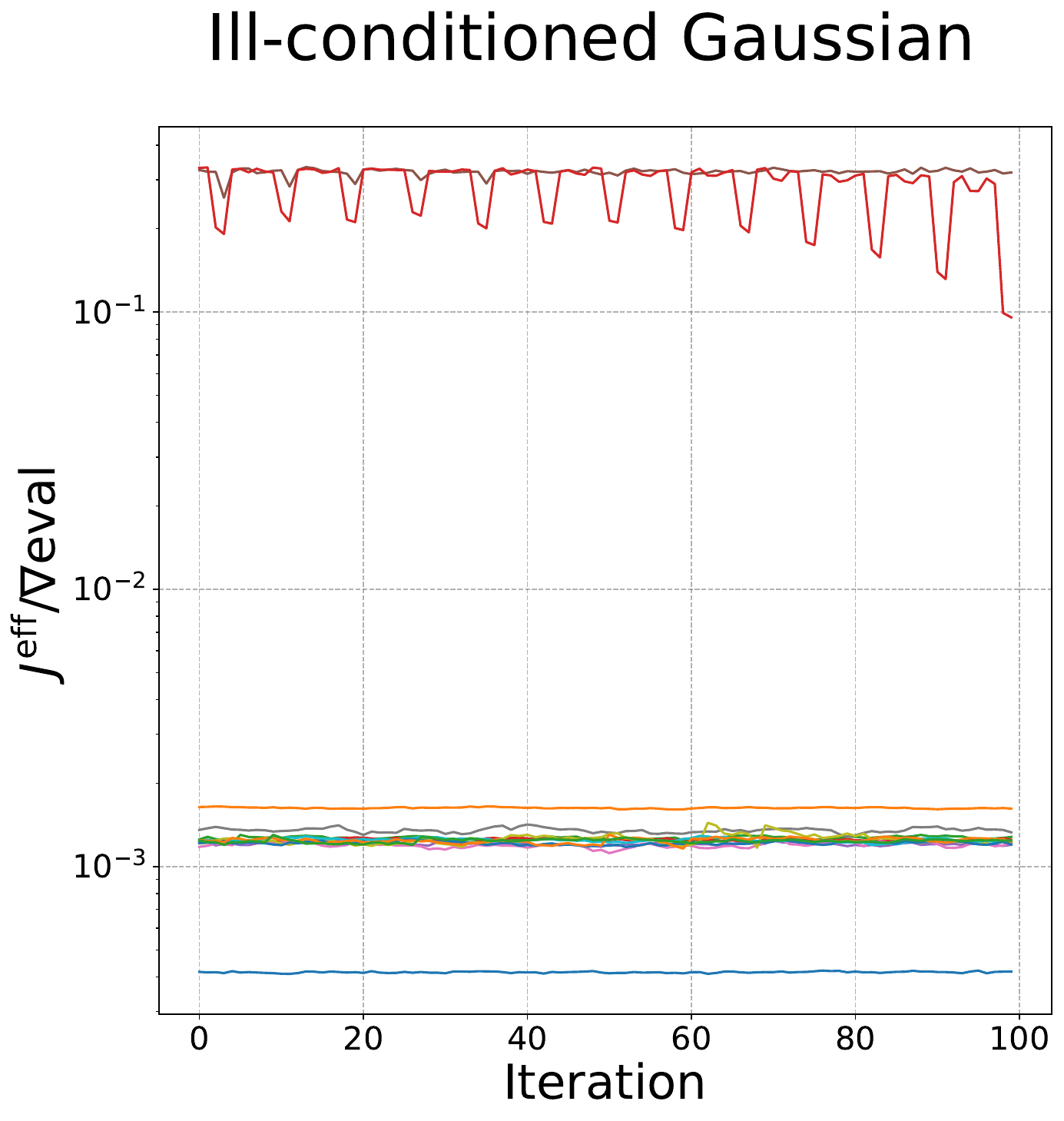}
    \end{subfigure}
    \hfill
    \begin{subfigure}{\SubFigureWidth\textwidth}
    \includegraphics[width=\textwidth]{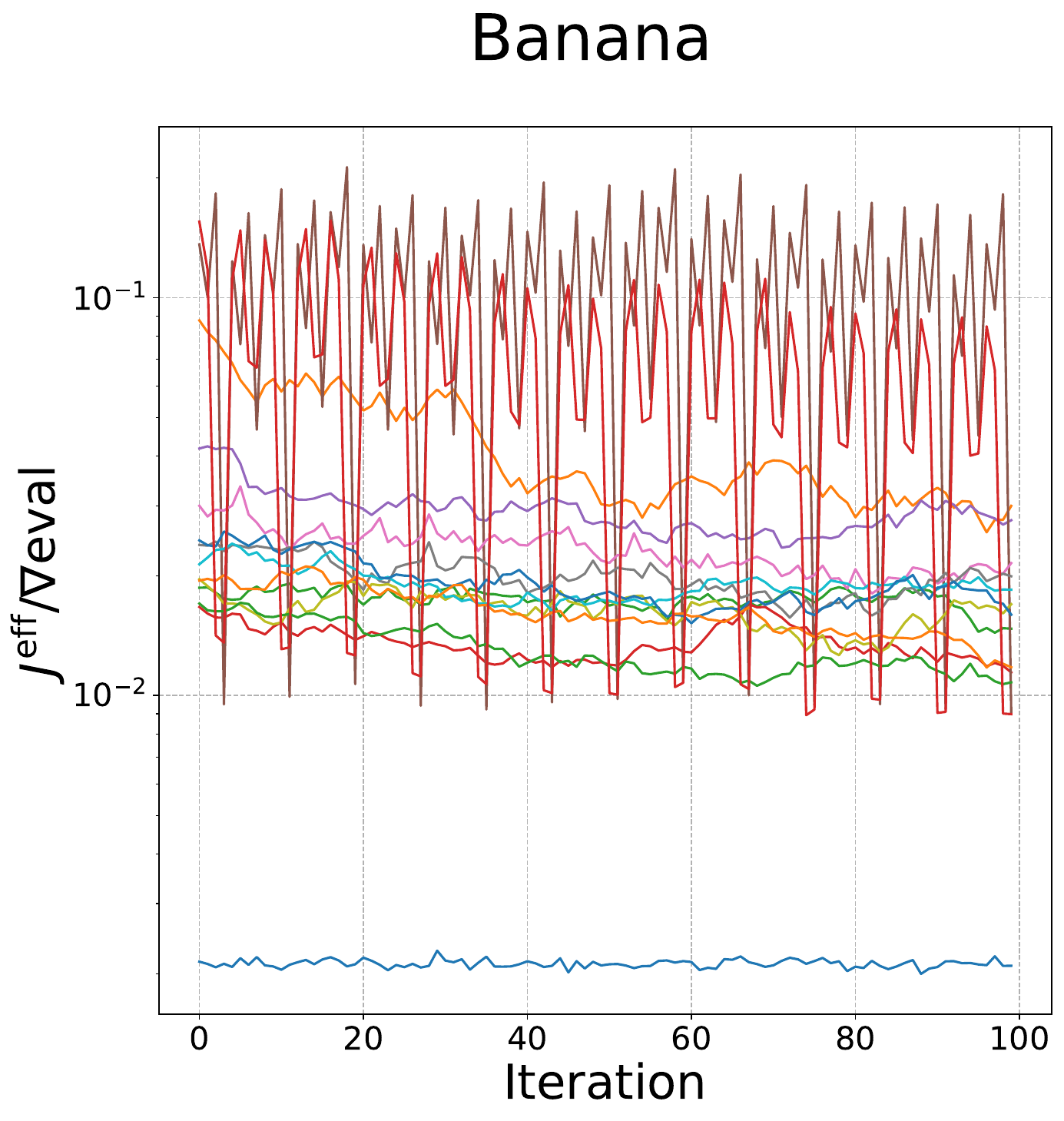}
    \end{subfigure}
    \hfill
    \begin{subfigure}{\SubFigureWidth\textwidth}
    \includegraphics[width=\textwidth]{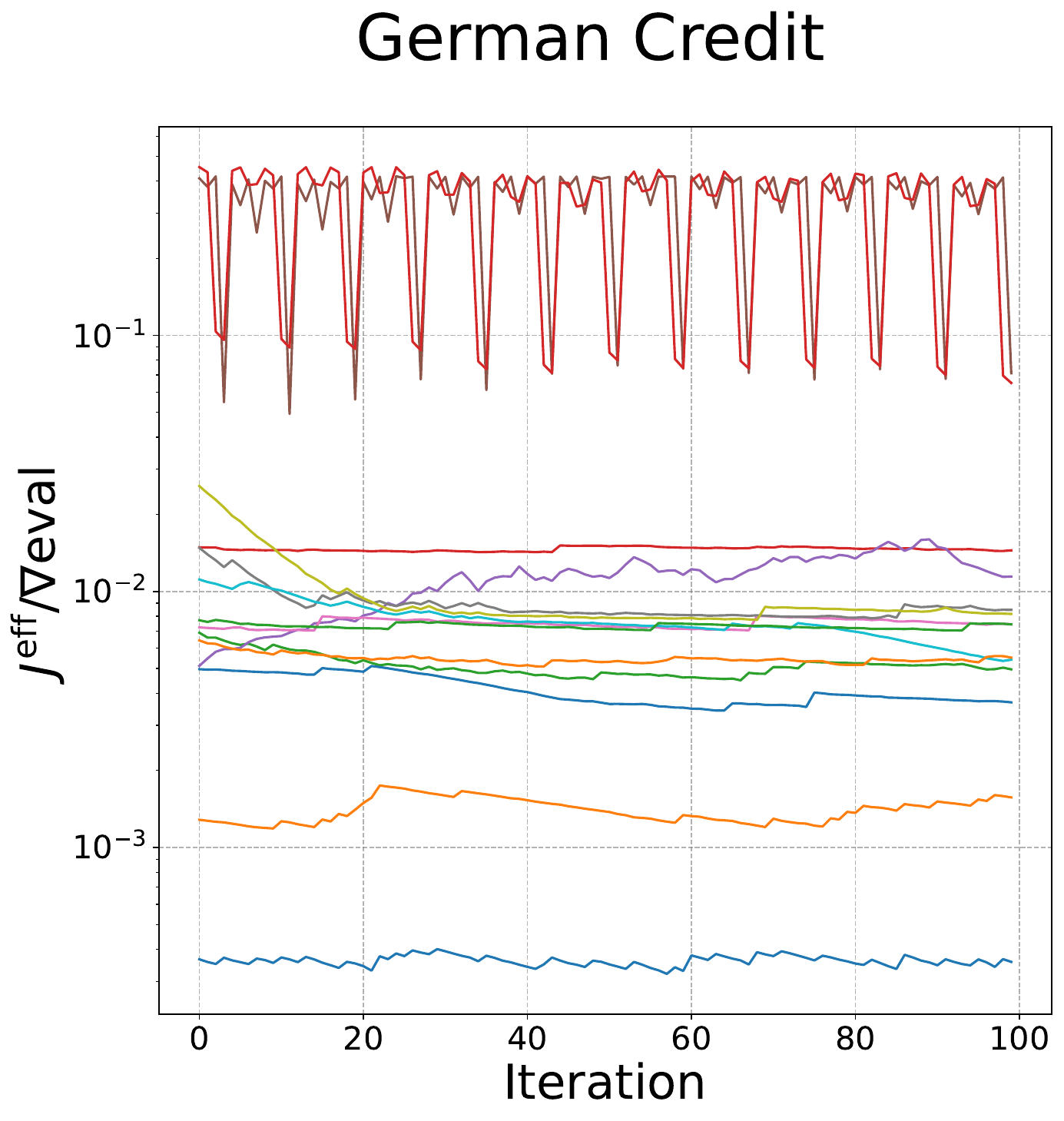}
    \end{subfigure}
    \caption{Number of effective samples per gradient evaluation per iteration for the four experiments.}
    \label{fig:neff_per_grad_eval} 
\end{figure*}

Fig. \ref{fig:neff_per_grad_eval} shows the number of effective samples per gradient evaluation ($J^\text{eff}/\nabla\text{eval}$) for each of the 4 experiments. It's clear to that  ChEES has a vastly greater $J^\text{eff}/\nabla\text{eval}$ than NUTS across all tasks with every RNG. This is reflected in Table~\ref{tab:grad_eval_ess} which shows the average number of gradient evaluations per sample and the average $J^\text{eff}/\nabla\text{eval}$ across iterations. For the Gaussian, Ill-conditioned Gaussian and Banana targets, NUTS had approximately 4 to 5 times more average $\nabla\text{eval}$ per particle than the ChEES methods and the same difference can be found in the average $J^\text{eff}/\nabla\text{eval}$. 

A far larger disparity in the number of $\nabla\text{eval}$ per particle and likewise in the $J^\text{eff}/\nabla\text{eval}$ can be seen in the German Credit task where NUTS used about 80 times more evaluations than some of the ChEES methods. In our implementation, the maximum tree depth of NUTS was set to $2^{11}$ meaning that no further leapfrog steps were taken beyond that number, even if the No-U-turn criterion in \eqref{eq:u-turn} was not met. NUTS took an average of $1952.19$ ($\nabla\text{eval}/N-1$) leapfrog steps as there is also a gradient calculation undertaken prior to the leapfrog steps. This is far greater than on any other task and the maximum tree depth limit was frequently reached during the sampling process. The ChEES proposals, are less effected by their use on real world data and don't see the same spike in $\nabla\text{eval}/N$. 

\begin{figure}[!t]
    \centering
    \newcommand{\SubFigureWidth}{0.23}
    \begin{subfigure}{\SubFigureWidth\textwidth}
    \includegraphics[width=\textwidth]{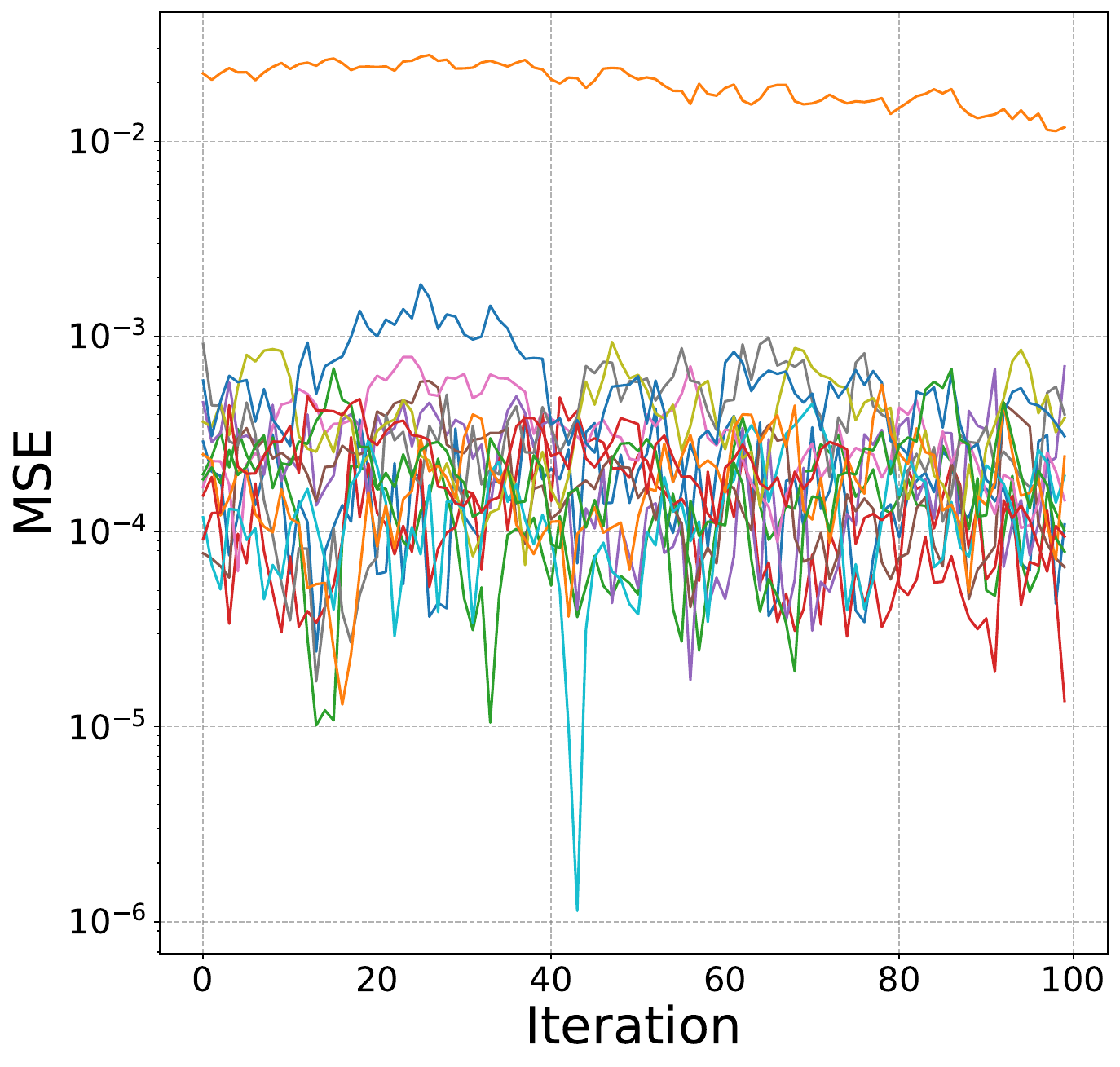}
    \caption{Mean MSE}
    \end{subfigure}
    \hfill
    \begin{subfigure}{\SubFigureWidth\textwidth}
    \includegraphics[width=\textwidth]{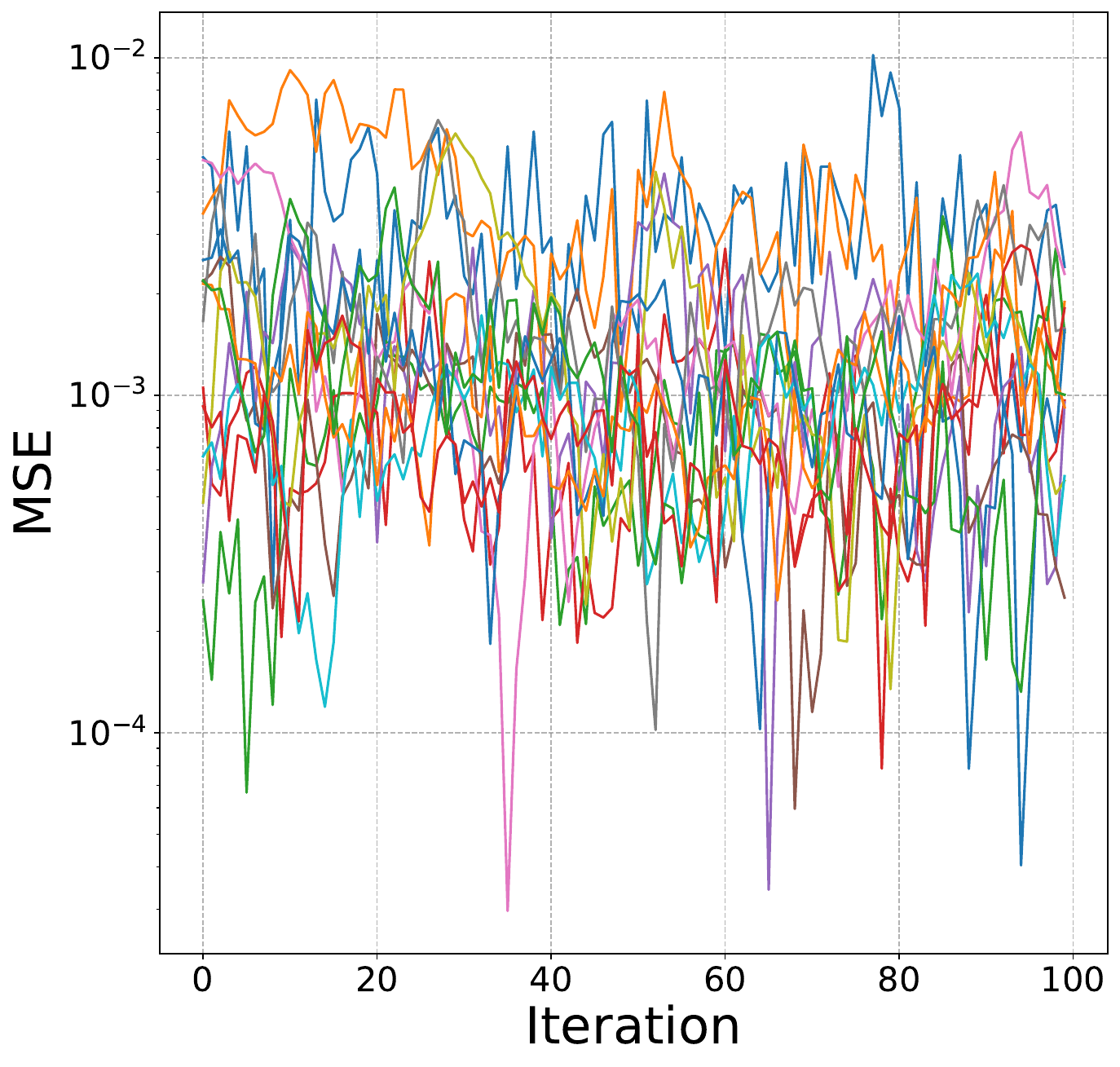}
    \caption{Variance MSE}
    \end{subfigure}
    \caption{MSE of the mean and variance estimates for the 5-dimensional Gaussian distribution obtained by the different proposals.}
    \label{fig:gauss_mses} 
\end{figure}

However, we do notice that on the ill-conditioned Gaussian example for nearly every method, we reach the maximum number of leapfrog steps manually set in the implementation. Therefore, we would need to increase the maximum trajectory length to in order to evaluate the effective sample size better. Despite reaching this limit though, all methods still produced good MSE on both the mean and variance showing that a longer trajectory length is potentially unnecessary. This also outlines that $J^\text{eff}/\nabla\text{eval}$ evaluation, although a good metric for understanding computational efficiency, should not be used as a singular metric as we notice this does not always translate to producing meaningful samples from the posterior. For example in the 5-D example \ref{fig:gauss_mses}, we see that the No Jitter method has a relatively good $J^\text{eff}/\nabla\text{eval}$, but a suboptimal MSE.  

The RNG that gives the best $J^\text{eff}/\nabla\text{eval}$ and MSE results overall seems to be the 1-d Halton method which was employed in the original ChEES \cite{hoffman2021adaptive} paper while the 1-d Sobol method also produces good results on the same metrics. However this again demonstrates the need to consider additional metrics as the performance of both RNGs on the Ill-conditioned Gauss and German Credit task are notably poorer. 

In Fig. \ref{fig:neff_per_grad_eval} the methods alternate for which has the highest $J^\text{eff}/\nabla\text{eval}$ across iterations and this is also reflected by Table \ref{tab:grad_eval_ess}. Our results support previous findings that jittering the trajectory length is helpful in drawing good samples from the posterior distribution \cite{neal2012mcmc}. 

\begin{figure}[!h]
    \centering
    \newcommand{\SubFigureWidth}{0.23}
    \begin{subfigure}{\SubFigureWidth\textwidth}
    \includegraphics[width=\textwidth]{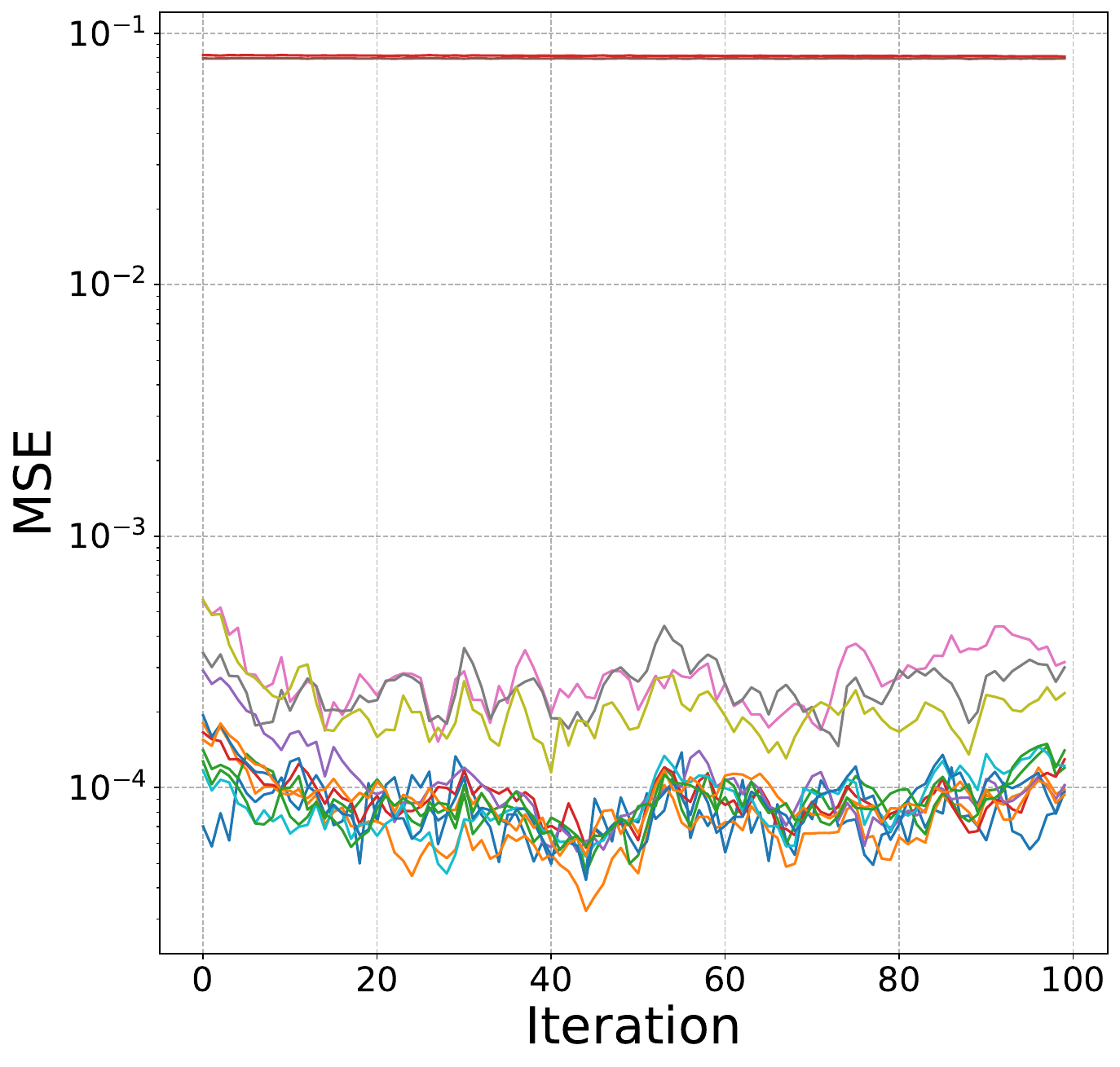}
    \caption{Mean MSE}
    \end{subfigure}
    \hfill
    \begin{subfigure}{\SubFigureWidth\textwidth}
    \includegraphics[width=\textwidth]{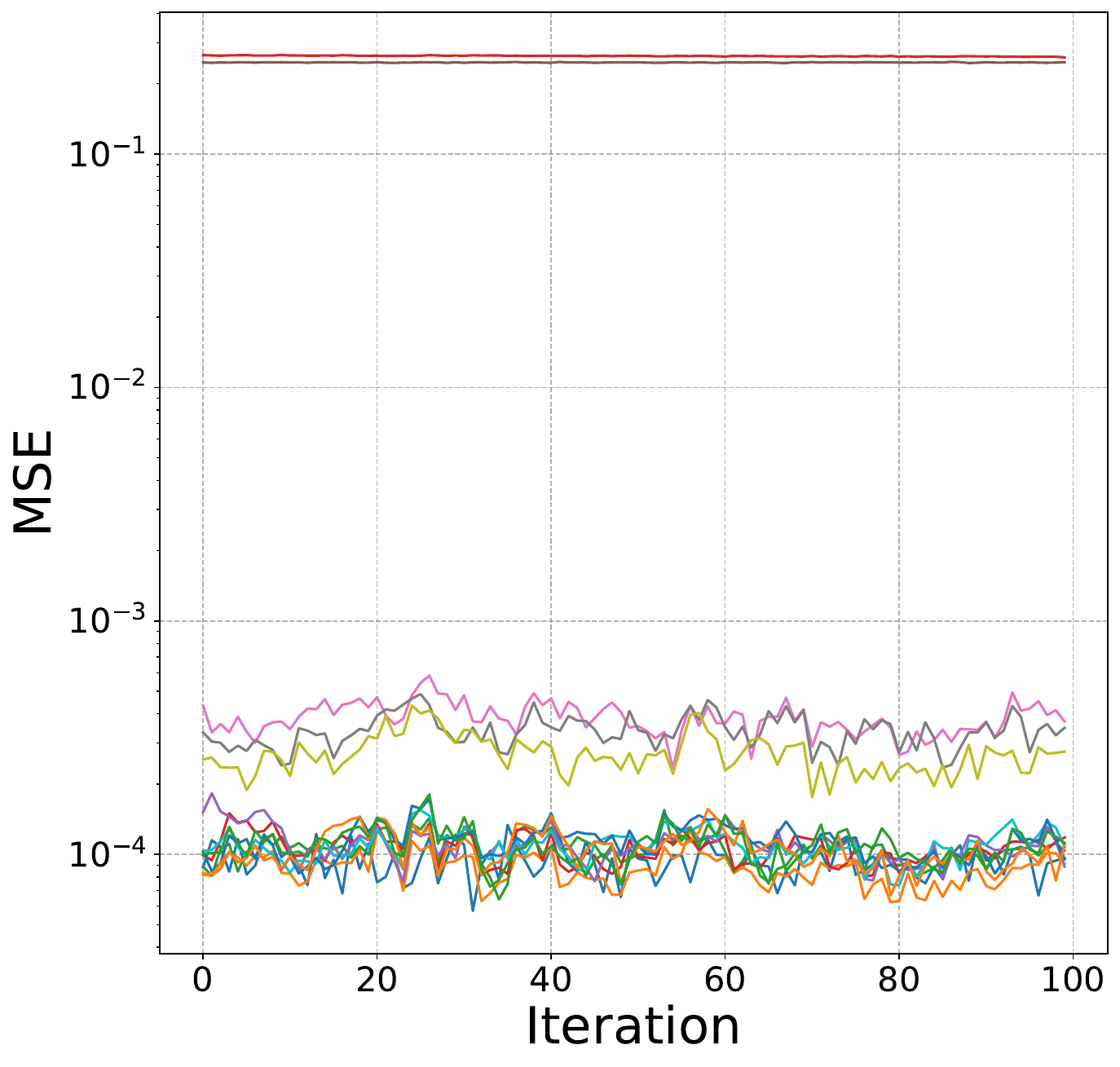}
    \caption{Variance MSE}
    \end{subfigure}
    \caption{MSE of the mean and variance estimates for the 100-dimensional Ill-conditioned Gaussian distribution obtained by the different proposals.}
    \label{fig:ic_gauss_mses} 
\end{figure}

\begin{figure*}[!ht]
    \begin{subfigure}{\textwidth}
        \includegraphics[width=\textwidth]{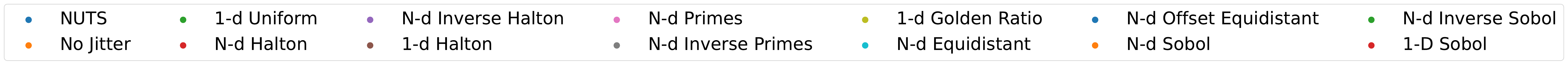}
    \end{subfigure}
    
    \includegraphics[width=\textwidth]{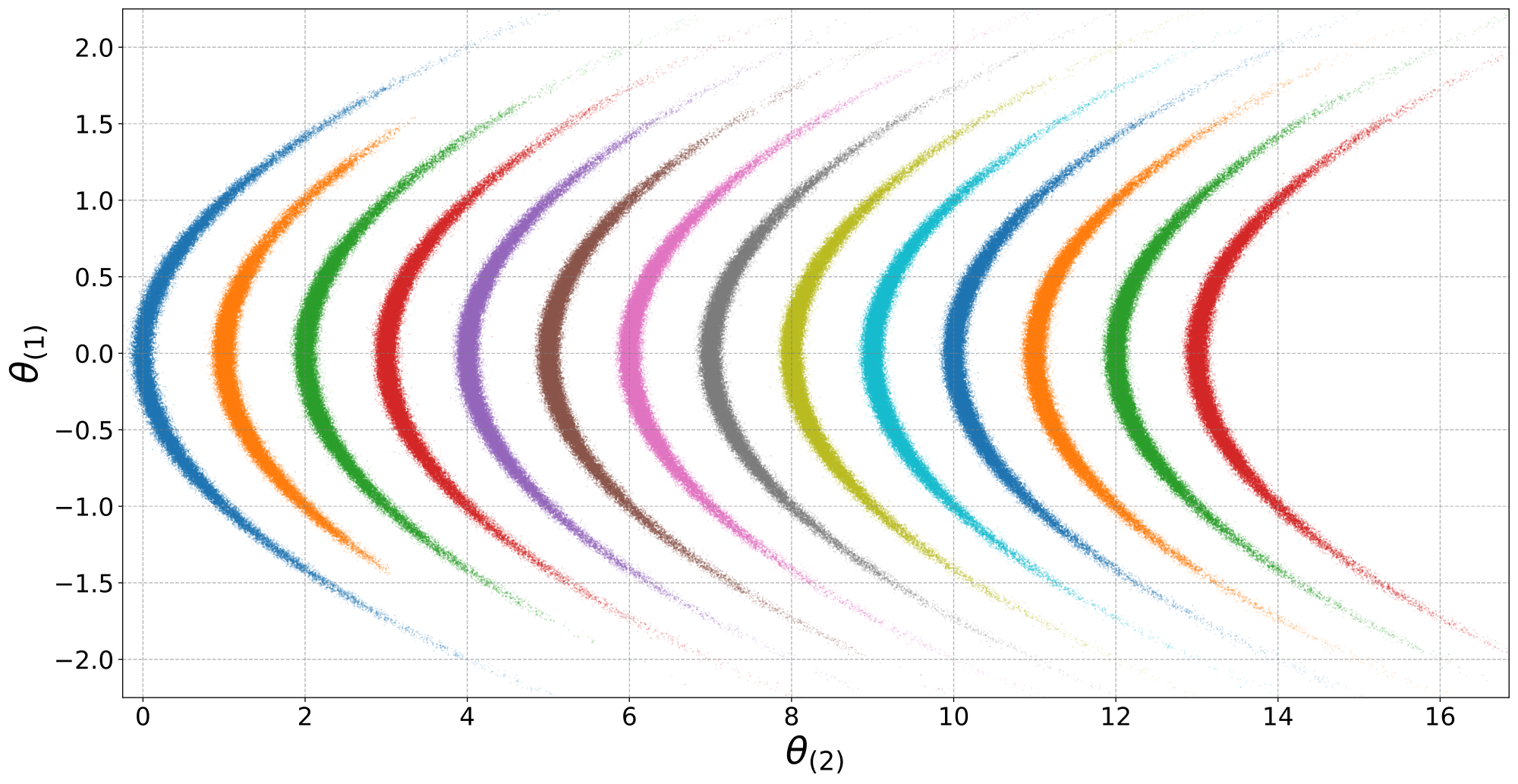}
    
    \caption{Position of all samples for the last 20 iterations of the banana distribution.}
    \label{fig:banana_samples} 
\end{figure*}

Fig. \ref{fig:gauss_mses} and \ref{fig:ic_gauss_mses} show the mean square error (MSE) for the mean and variance of the Gaussian and Ill-conditioned Gaussian. The mean and variance are realised using \eqref{eq:realised_estimates} and an average is then taken over the dimensions. NUTS is clearly the first to converge and the ChEES proposals converge to a similar MSE to NUTS within 15 iterations. In the Banana distribution NUTS explores furthest into the tails as shown in Fig. \ref{fig:banana_samples} which shows the positions of all particles across the last twenty iterations. Once again NUTS is the best proposal on the German Credit logistic regression task but the difference to the ChEES methods is marginal and therefore we may see another method produce marginally better results with more/different starting seeds. 

\begin{table*}[!t]
    \centering
    \begin{tabular}{|c|c|c|c|c|c|c|}
        \hline
         &
         Accuracy & Precision & Recall & F1 Score & Specificity & AUROC \\
        \hline
        \textbf{NUTS} & 0.78 & 0.66 & 0.53 & 0.58 & 0.89 & 0.71 \\
\textbf{No Jitter} & 0.78 & 0.66 & 0.53 & 0.58 & 0.89 & 0.71 \\
\textbf{1-d Uniform} & 0.76 & 0.64 & 0.42 & 0.51 & 0.90 & 0.66 \\
\textbf{N-d Halton} & 0.76 & 0.63 & 0.46 & 0.53 & 0.89 & 0.67 \\
\textbf{N-d Inverse Halton} & 0.77 & 0.65 & 0.44 & 0.53 & 0.90 & 0.67 \\
\textbf{1-d Halton} & 0.76 & 0.65 & 0.41 & 0.50 & 0.91 & 0.66 \\
\textbf{N-d Primes} & 0.77 & 0.65 & 0.44 & 0.53 & 0.90 & 0.67 \\
\textbf{N-d Inverse Primes} & 0.78 & 0.68 & 0.47 & 0.56 & 0.91 & 0.69 \\
\textbf{1-d Golden Ratio} & 0.77 & 0.64 & 0.47 & 0.54 & 0.89 & 0.68 \\
\textbf{N-d Equidistant} & 0.76 & 0.64 & 0.42 & 0.51 & 0.90 & 0.66 \\
\textbf{N-d Offset Equidistant} & 0.77 & 0.66 & 0.46 & 0.54 & 0.90 & 0.68 \\
\textbf{N-d Sobol} & 0.77 & 0.65 & 0.44 & 0.53 & 0.90 & 0.67 \\
\textbf{N-d Inverse Sobol} & 0.77 & 0.65 & 0.44 & 0.53 & 0.90 & 0.67 \\
\textbf{1-D Sobol} & 0.76 & 0.67 & 0.37 & 0.48 & 0.92 & 0.65 \\
        \hline
    \end{tabular}
    \caption{Logistic Regression Results for the German Credit Dataset (larger is better).}
    \label{tab:gc_metrics}
\end{table*}

%% file: 6_conc.tex
\section{Conclusions}\label{sec:conclusion}
In this paper we incorporate ChEES as a proposal into an SMC framework with a change of variables L-kernel and explore the use of different RNG options to jitter the trajectory length. We demonstrate that ChEES is far more efficient in terms of the number of effective samples it achieves per gradient evaluation with this difference being particularly pronounced on the real-world German Credit logistic regression and Banana distribution examples. 

We have also investigated the different RNG methods for jittering outperform no-jittering within the ChEES algorithm with the best overall performance coming from the 1-d Halton and 1-d Sobol methods. We note that the depth has been reached for many methods on the Ill-conditioned Gaussian example which requires more investigation. Despite hitting the maximum number of leapfrog steps, the MSE performance is still on par with NUTS. 

Further work could focus on including step size adaption methods (such as dual averaging) into SMC-ChEES to minimise the number of hyper-parameters that need to be manually tuned. As SMC-ChEES picks a singular trajectory length which it then jitters, it is competitive with NUTS in scenarios where NUTS trajectory lengths may not differ significantly. Therefore, it would be useful to investigate ChEES on example problems where this is not the case, such as the funnel distribution \cite{neal2003slice}. 

SMC-ChEES could also be further evaluated on more real-world data to see if the benefit of using alternative RNG methods with ChEES persists across other tasks.